# Data-driven Learning of Probabilistic Model of Binary Droplet Collision for Spray Simulation

Weiming Xu[1,2#], Tao Yang[1,2#], and Peng Zhang[1,2*]

1. Department of Mechanical Engineering, City University of Hong Kong, Kowloon Tong, Kowloon, 999077, Hong Kong
2. Shenzhen Research Institute, City University of Hong Kong, Shenzhen, 518057, P. R. China

**Abstract**

Binary droplet collisions are ubiquitous in dense sprays. Traditional deterministic models cannot adequately represent transitional and stochastic behaviors of binary droplet collision. To bridge this gap, we developed a probabilistic model by using a machine learning approach, the Light Gradient-Boosting Machine (LightGBM). The model was trained on a comprehensive dataset of 33,540 experimental cases covering eight collision regimes across broad ranges of Weber number, Ohnesorge number, impact parameter, size ratio, and ambient pressure. The resulting machine learning classifier captures highly nonlinear regime boundaries with 99.2% accuracy and retains sensitivity in transitional regions. To facilitate its implementation in spray simulation, the model was translated into a probabilistic form, a multinomial logistic regression, which preserves 93.2% accuracy and maps continuous inter-regime transitions. A biased-dice sampling mechanism then converts these probabilities into definite yet stochastic outcomes. This work presents the first probabilistic, high-dimensional droplet collision model derived from experimental data, offering a physically consistent, comprehensive, and user-friendly solution for spray simulation.

**Keywords**: Spray simulation; Droplet collision; Machine learning; LightGBM algorithm; Logistic regression modeling; Stochastic classification

---

[*] Corresponding author
  E-mail address: penzhang@cityu.edu.hk, Tel: (852)34429561.

[#] The authors equally contributed to the work.

# 1. Introduction

Collision of binary droplets, as a common phenomenon in nature and technology, underpins key processes in raindrop formation McFarquhar (2004), inkjet printing van der Bos, et al. (2014), drug delivery Kichatov, et al. (2024), and various spraying processes Sirignano (1983), Kim, et al. (2009), Makhnenko, et al. (2021). A prominent example is its role in combustion engines, where the outcome of droplet collisions governs droplet size in fuel atomization, thereby determining overall combustion efficiency and pollutant emissions Ren, et al. (2019), An, et al. (2024), Li, et al. (2024). Similarly, these collisions influence cloud microphysics and precipitation patterns in meteorology Hu (1995), while in pharmaceutical manufacturing, they affect the uniformity of spray-dried particles for drug delivery systems Mishra, et al. (2023). In spray simulations, modeling droplet collisions is critical for capturing droplet size and velocity distributions, overall atomization behaviors, and subsequent processes like evaporation and combustion. For example, the Eulerian-Lagrangian approach is widely used to track individual droplet trajectories within a continuous gas phase, where the precision in predicting collision outcomes directly dictates the reliability of simulating key processes such as mass exchange, spray evolution, and heat transfer Lain and Sommerfeld (2020), Sazhin (2022).

As shown in Fig. 1, extensive studies Jiang, et al. (1992), Qian and Law (1997), Pan, et al. (2009), Tang, et al. (2012), Huang, et al. (2019), Pan, et al. (2019), Huang and Pan (2021), Xia, et al. (2025), Zhou, et al. (2022) over the past decades has identified the regimes of binary droplet collision outcomes, such as (I) soft coalescence, (II) bouncing, (III) hard coalescence, (IV) reflexive separation, (V) stretching separation, (VI) rotational separation, (VII) finger separation, and (VIII) splashing, across various liquids and ambient pressures. For the collision of the same liquid droplets at an atmospheric pressure $p_0$, these outcomes are governed by key dimensionless parameters, such as the Weber number $We = \rho D_s U^2/\sigma$, Ohnesorge number $Oh = \mu/(\rho D_s \sigma)^{1/2}$, impact parameter $B = 2\chi/(D_s + D_l)$, and size ratio $\Delta = D_l/D_s$, where $\rho$ is the droplet density, $\mu$ the viscosity, $\sigma$ the surface tension, $D_s$ and $D_l$ the diameters of the small and large droplets, $U$ the relative velocity, and $\chi$ the offset distance between the droplet centers perpendicular to the relative velocity.

These regime boundaries are important for developing collision models. As such, previous researchers formulate these composite collision models by integrating specific combinations of the critical boundaries between different outcomes based on analytical or empirical correlations Jiang, et al. (1992), Al-Dirawi and Bayly (2019), Sui, et al. (2023), Sommerfeld and Kuschel (2016), Ashgriz and Poo (1990), Suo and Jia (2020). The composite models Kim, et al. (2009), Georjon and Reitz (1999), Post and Abraham (2002), Luret, et al. (2010), Tyurenkova, et al. (2024) are based on a combination of theoretical analysis (e.g., energy conservation and viscous dissipation), experimental observations, and numerical simulations. Around 3–4 critical boundaries are identified to separate several main regimes in multi-parameter nomograms, such as the lower bouncing boundary Estrade, et al. (1999), the reflexive separation-coalescence boundary Ashgriz and Poo (1990), and the stretching separation-coalescence boundary Brazier-Smith, et al. (1972).

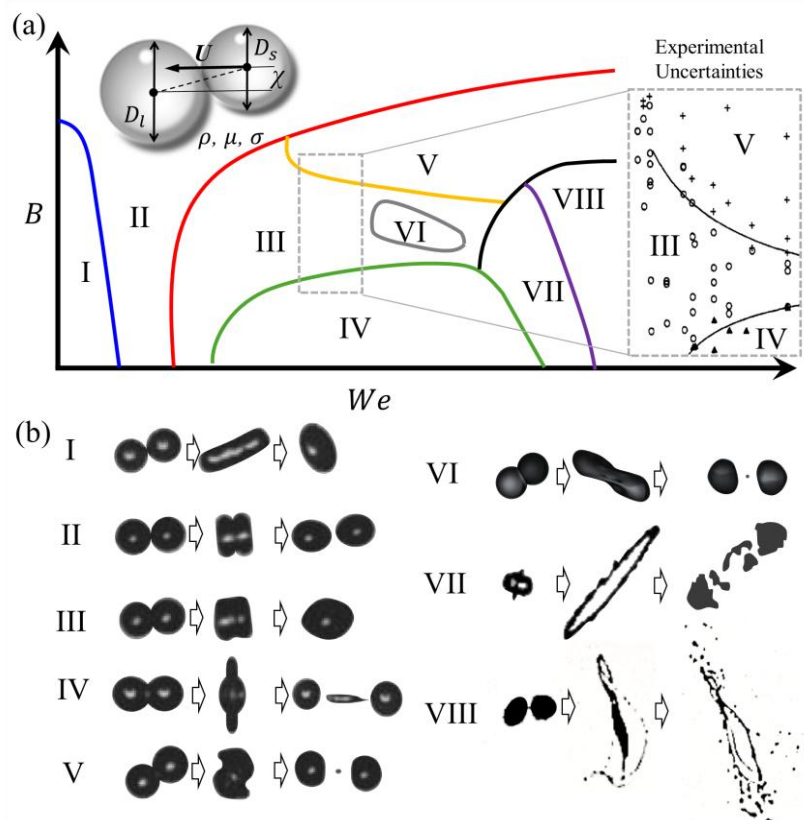

Fig.1 Binary droplet collision in gaseous environment: (a) Schematic regime nomogram in the $We-B$ parameter space, the inserted figure from the experiment results in Fig. 2 of Aashgriz and Poo Ashgriz and Poo (1990); (b) typical outcomes from the experiments of Qian and Law Qian and Law (1997), Pan et al. Pan, et al. (2019), and Zhou et al. Zhou, et al. (2022), namely I (soft coalescence after minor

deformation), II (bouncing), III (hard coalescence after substantial deformation), IV (reflexive separation, near head-on separation), V (stretching separation, off-centre separation), VI (rotational separation), VII (finger separation), and VIII (splashing). Note that Regime I occurs only in a very narrow low-Weber-number region; its detailed visualization is provided in the Supplementary Material.

Previous models often fall short due to their limited consideration of collision outcomes and dimensionless parameters for comprehensive regime boundaries Agarwal, et al. (2019), Yu and Chang (2026). Using the composite model by Munnannur and Reitz [36], Agarwal et al. Agarwal, et al. (2019) assessed its accuracy at 64% against earlier datasets Qian and Law (1997), Ashgriz and Poo (1990), Estrade, et al. (1999) and 43% against a recent experiment Sommerfeld and Kuschel (2016). These shortcomings underscore the significant need for advanced modelling approaches that integrate comprehensive experimental data and sophisticated techniques to enhance predictive fidelity. Furthermore, the existing experimental data are not uniformly distributed in the parameter ranges. For example, the collisions with vanishingly small Weber numbers (corresponding to head-on collisions) are relatively rare compared with great many non-zero Weber numbers. Consequently, the previous models of head-on collision often relied on the data extrapolation of small Weber numbers to zero Weber number. In addition, experimental uncertainties were largely neglected in previous models; although deterministic regime curves may appear well separated in selected datasets, repeated experiments under nominally identical conditions can yield inconsistent or even contradictory outcomes in certain transitional parameter ranges. All of these observations of the exising models necessitate new collision models that account for the paucity, uncertainty, and high-dimension of experimental data.

In recent years, machine learning (ML) has sparked growing interest in droplet dynamics, driven by the convergence of enhanced computational resources, high-fidelity experimental datasets, and mature ML frameworks. Different ML models have been adopted to predict geometrical properties or physical behaviors observed in droplet flows, such as droplet spreading on a surface Au-Yeung and Tsai (2023), Tang, et al. (2023), droplet splashing on a surface Yee, et al. (2023), Pierzyna, et al. (2021), Ye, et al. (2023), and binary droplet collision modeling Yu and Chang (2026), Agarwal (2021), [43]. These ML-based methods offer a compelling pathway to overcome the limitations of traditional empirical models in numerical simulations, providing a means to enhance predictive accuracy while achieving significant gains in computational efficiency Kochkov, et al. (2021), Salehi, et al. (2025).

Agarwal Agarwal (2021) collected 7898 experimental data points and selected five base features ($We$, $B$, $\Delta$, $P$, and $\mu$) and three physical-model-based curves as hyperparameters for optimizing non-linear ML classifiers, including decision trees, random forests, K-nearest neighbors, and neural networks. Those ML classifiers, with accuracies over 90%, significantly outperform the physics-based models Estrade, et al. (1999), Munnannur and Reitz (2007) to predict four regimes (coalescence, bouncing, stretching separation, and reflexive separation). Subsequently, Yu and Chang Yu and Chang (2026) extended the training dataset into 30809 experimental data points and integrated an analytical composite collision model into the XGBoost algorithm for 94% accuracy prediction. However, a significant research gap exists: a highly accurate and usable ML model for droplet collision outcomes in spray simulations, particularly one based on comprehensive experimental data and more sophisticated techniques, has yet to be established. Existing data-driven studies for droplet collision modeling are predominantly limited to deterministic or standard ML classification approaches, typically considering a reduced number of regimes and relatively narrow parameter ranges Estrade, et al. (1999), Yu and Chang (2026), Hu, et al. (2017), Sui, et al. (2019). As a result, transitional and stochastic collision behaviors are largely neglected.

In this study, we aim to develop a robust and highly accurate machine learning-based droplet collision model using a substantial and comprehensive database. To the best of our knowledge, this is the first framework that integrates probabilistic regime prediction with fuzzy decision boundaries, explicit regression-based probability modeling, and biased-dice sampling in droplet collision modeling for spray simulations. The data points, which cover eight regimes from I to VIII within the five parameter spaces of $We$, $Oh$, $B$, $\Delta$, and $P = p/p_0$, are collected from previous experiments. The proposed ML approach employs the Light Gradient Boosting Machine (LightGBM) algorithm, which utilizes Gradient-based One-Side Sampling (GOSS) for accelerated training without loss of accuracy, while employing Exclusive Feature Bundling (EFB) for the effective handling of sparse features. To facilitate the implementation of the new model in spray simulations, the method expresses the regime boundaries through probabilistic formulations captured by multinomial logistic regression, then employs a biased-dice sampling mechanism to translate these fuzzy probabilities into single realizations, thereby providing simulation-ready outcomes for user-friendly integration into numerical computations.

## 2. Methodology
### 2.1 Data Collection and Features

Following a Buckingham – Pi dimensional analysis [48], five independent dimensionless parameters ($We$, $Oh$, $\Delta$, $B$, and $P$) are identified as a minimal and physically consistent set governing binary droplet collision outcomes. A comprehensive database of 33,540 experimental data points was compiled from 26 previous studies Kim, et al. (2009), Jiang, et al. (1992), Qian and Law (1997), Tang, et al. (2012), Huang, et al. (2019), Pan, et al. (2019), Huang and Pan (2021), Xia, et al. (2025), Zhou, et al. (2022), Al-Dirawi and Bayly (2019), Sui, et al. (2023), Sommerfeld and Kuschel (2016), Brenn, et al. (2001), Brenn and Kolobaric (2006), Rabe, et al. (2010), Planchette, et al. (2010), Foissac, et al. (2010), Kuschel and Sommerfeld (2013), Hinterbichler, et al. (2015), Pan, et al. (2016), Reitter, et al. (2017), Finotello, et al. (2018), Sommerfeld and Pasternak (2019), Sommerfeld and Pasternak (2021), McCarthy, et al. (2022), Baumgartner, et al. (2022), all of which concern Newtonian fluids. The data cover eight collision regimes: regime I (203 points), II (8,785), III (13,275), IV (2,518), V (8,162), VI (259), VII (83), and VIII (255) in the parameter ranges: $We = 0\sim2000$, $Oh = 2.7\times10^{-3}\sim5.5\times10^{-1}$, $B = 0\sim1$, $\Delta = 1\sim5$, and $P = 0.6\sim9.0$. In brief, we present the features of the collected dataset. The $We - B$ nomogram reveals an overall distribution of experimental data points across eight collision regimes in Fig. 2(a). The data distribution of each regime is identified through a kernel smoothing function estimate.

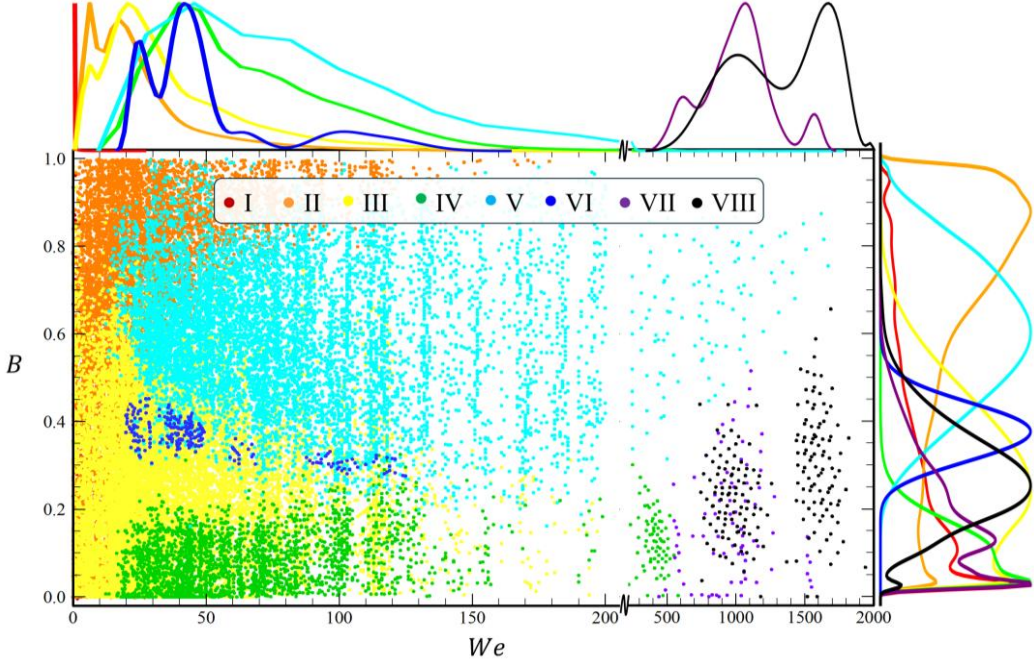

Fig 2. The complete dataset of eight droplet collision regimes in the $We - B$ space, including I (soft coalescence after minor deformation), II (bouncing), III (hard coalescence after substantial deformation), IV (reflexive separation, near head-on separation), V (stretching separation, off-centre separation), VI (rotational separation), VII (finger separation), and VIII (splashing). The curves represent the data distribution of each regime via estimating point density along univariate $We$ and $B$.

It is noted that the data points are densely clustered in the low-Weber number region (We < 100), primarily populating regimes I -VI. These points of regime I are concentrated at low $B$ and low $We$ ($< 5$), while those of Regime II extend toward higher $B$ values with higher $We$. As $We$ increases toward 200, the distribution of regime III narrows gradually and forms a central band around $B = 0 - 0.4$, bounded by the reflexive separation-coalescence and stretching separation-coalescence lines. Higher-$B$ regime IV is $50 < We < 200$, while lower-$B$ regime V is $20 < We < 150$. Regime VI emerges prominently in a small region of $B$ ($\approx 0.4$) and $We$ (20~120). Beyond $We \approx 200$, regimes VII–VIII exhibit sparser in the range of $B = 0$~0.6.

Multi-dimensional parameter spaces in Fig. 3 further elucidate how the $We$ and $B$ modulate regime distributions through interactions with $Oh$, $P$, and $\Delta$. In Fig. 3(a), this $We - Oh$ regime diagram reveals that the distribution of collision regimes, where outcomes are influenced by a balance between inertial and viscous forces. At very low $We$, Regime I dominates regardless of $Oh$. As $We$ increases, inertial forces drive increasingly disruptive outcomes, such as regimes IV and V. In Fig. 3(b), the collision experiments were conducted in a wide range of $We$ at $P = 1$, while cases with higher $P$ rarely appear at high $We$ (>50).

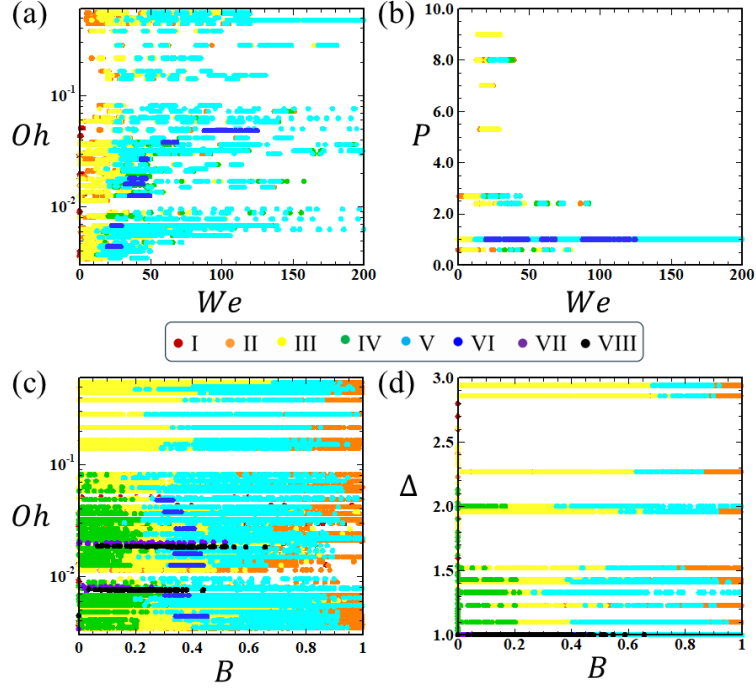

Fig 3. The dataset of eight droplet collision regimes in the two-dimensional parametric spaces of (a) $We - Oh$, (b) $We - P$, (c) $B - Oh$, and (d) $B - \Delta$. The collision regimes include I (soft coalescence after minor deformation), II (bouncing), III (hard coalescence after substantial deformation), IV (reflexive separation, near head-on separation), V (stretching separation, off-centre separation), VI (rotational separation), VII (finger separation), and VIII (splashing).

In Fig. 3(c), the results demonstrate that the collision regime is highly relative to both droplet viscosity and the alignment of the collision. The droplet outcomes of bouncing, hard coalescence, and stretching separation all exist in the whole range of $Oh = 2.7 \times 10^{-3} \sim 5.5 \times 10^{-1}$, while other ones occur at $Oh < 0.1$. Figure 3(d) shows the effects of droplet size disparity and collision alignment on regimes. The size ratio in previous studies is mainly focused on the range of 1.0-1.5, and high $\Delta$ values are rare.

**2.2 Data-driven Learning Articulture**

Reliable regime identification is critical for predictive modeling of droplet interactions in spray simulations. However, conventional expert-defined regime maps are limited by empirical assumptions and rigid boundaries Al-Dirawi and Bayly (2019),

Sui, et al. (2023), while purely data-driven machine learning models often suffer from black-box behavior that hinders physical interpretation and analytical use. To overcome these limitations, we developed a hybrid modeling strategy that extracts an explicit and probabilistic model from experimental data, providing both accuracy and transparency for practical simulations, as shown in Fig. 4.

To learn nonlinear relations among impact parameters, a Light Gradient Boosting Machine (LightGBM) was employed to capture the stochastic nature of regime transitions beyond deterministic classifications Ke, et al. (2017). The probabilistic outputs naturally describe smooth transitions between regimes and encode the inherent variability observed in experiments. Here, we outline the data-driven articulture for identifying the collision regimes and regressing a set of correlations of eight collision outcomes in the five dimension parameter space. Specifically, the learned probability fields were projected onto a compact analytical form through multinomial logistic regression. This step transforms the high-dimensional decision-tree predictions into explicit analytical expressions, facilitating both interpretation and visualization of fuzzy boundaries within reduced parameter subspaces. The final stage introduces a stochastic sampling procedure, here formulated as a biased dice, in which the analytically regressed probabilities determine the likelihood of each outcome. In contrast to a single deterministic label, this representation reproduces the experimentally observed variability in transition regions while remaining a user-friendly application in large-scale simulations.

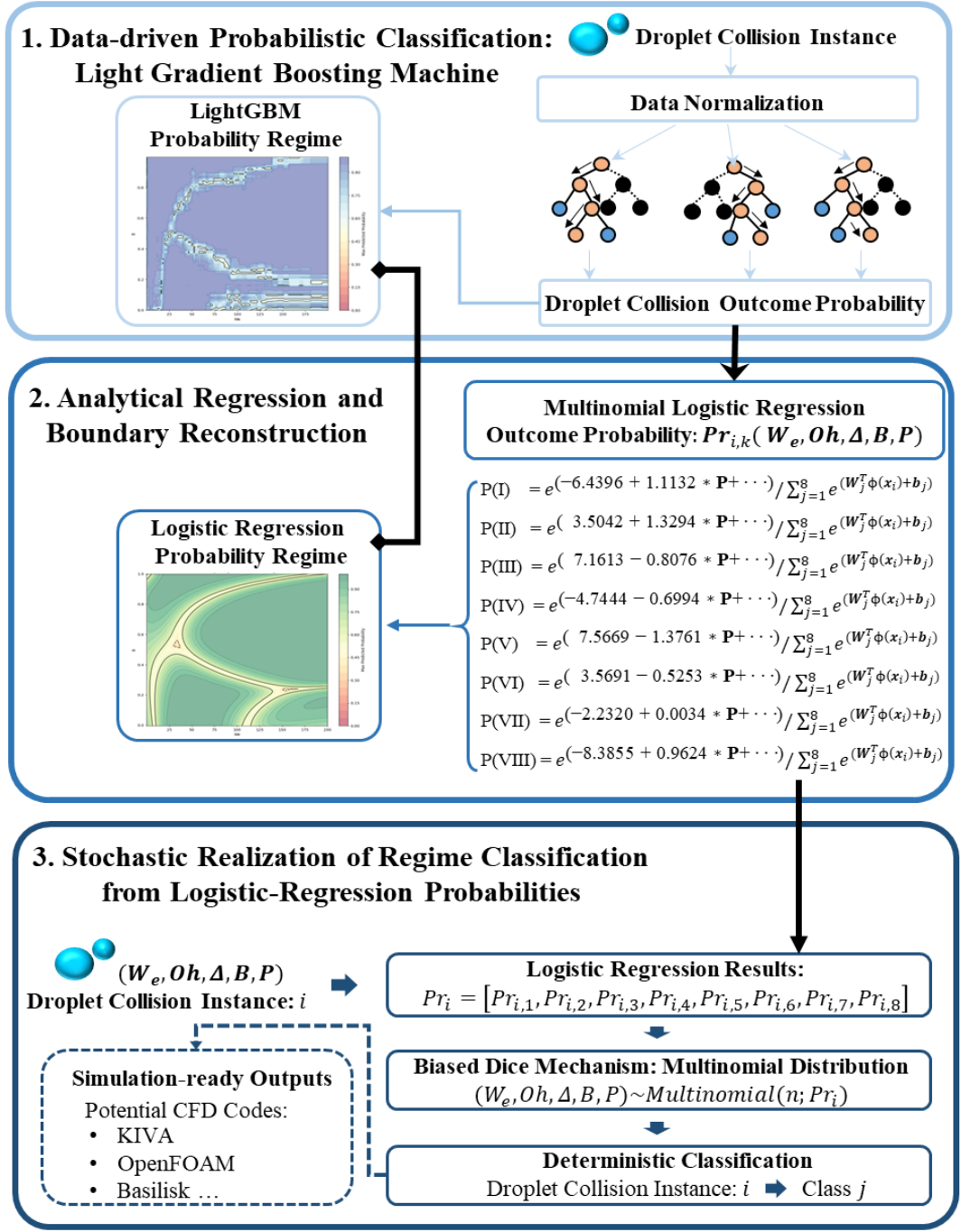

Fig 4. Schematic of the integrated machine learning workflow for droplet collision prediction, including data-driven probabilistic classification, analytical regression and boundary reconstruction, and stochastic realization of regime classification from logistic-regression probabilities

### 2.2.1 Theoretical Underpinnings of LightGBM Algorithm

LightGBM Ke, et al. (2017), an advanced implementation of gradient boosting decision trees, offers distinct advantages in addressing these challenges. Its capacity to handle large, imbalanced, high-dimensional datasets makes it particularly suitable for modeling the complex structure of droplet collision regimes. Accurate classification of droplet collision regimes requires a model capable of resolving highly nonlinear boundaries within a high-dimensional parameter space, while maintaining sensitivity to rare events in imbalanced data. Near regime transitions, the stochastic nature of droplet collision outcomes also challenges the applicability of traditional hard boundaries, making a fuzzy boundary formulation a more physically consistent and informative approach.

As shown in Fig. 5, each input sample $(\boldsymbol{x}_i, \boldsymbol{y}_i)$ is standardized by Z-score normalization Henderi, et al. (2021), and the model estimates the probability of class $k$ as

$$\boldsymbol{p}_{i,k} = \frac{e^{\hat{z}_{i,k}}}{\sum_{j=1}^{K} e^{\hat{z}_{i,j}}} \tag{1}$$

where $\boldsymbol{x}_i \in \mathbb{R}^5$ denotes the collision parameters, $\boldsymbol{y}_i \in \{1,2,3,\cdots K\}$ represents the observed outcomes, and $\hat{\boldsymbol{z}}$ represents the prediction logit score produced by the classifier, measuring the model's confidence in each class. The softmax transformation ensures non-negativity and normalization of probabilities, naturally forming the fuzzy decision boundary and continuous probability representations.

This method further enhances computational efficiency through several algorithmic innovations. Histogram-based binning discretizes continuous features to significantly reduce the cost of split evaluation. Gradient-based One-Side Sampling (GOSS) allocates greater computational weight to samples with larger gradients, thereby concentrating learning on the most informative instances while preserving predictive accuracy. Exclusive feature bundling (EFB) addresses high-dimensional sparsity by grouping mutually exclusive features, reducing memory demand without degrading performance. Leaf-wise tree growth prioritizes splits that yield the largest gain, enabling deeper trees with fewer splits and finer resolution of decision boundaries. The specific mathematical details are discussed in the Supplementary Material.

Within the present work, the trained LightGBM classifier produces a set of class probabilities $p_{i,k}$, which encode the structure of fuzzy boundaries in the high-dimensional parameter space of binary droplet collisions. These probabilities encapsulate not only the classification outcome but also the inherent uncertainty near transitional regimes. As such, they serve as the essential quantitative foundation for the subsequent analytical regression and biased-dice sampling stages, enabling a physically consistent and computationally efficient representation of collision regime dynamics.

A grid search was performed over the following ranges: learning rate [0.01, 0.1, 0.2], maximum tree depth [3, 5, 10], number of boosting iterations [50, 100, 200], and number of leaves per tree [31, 50, 100]. The final chosen parameters—learning rate of 0.2, a maximum tree depth of 10, 200 boosting iterations, and 31 leaves per tree—achieved a best cross-validation F1-score of 0.921, indicating both strong predictive performance and robustness across all collision regimes.

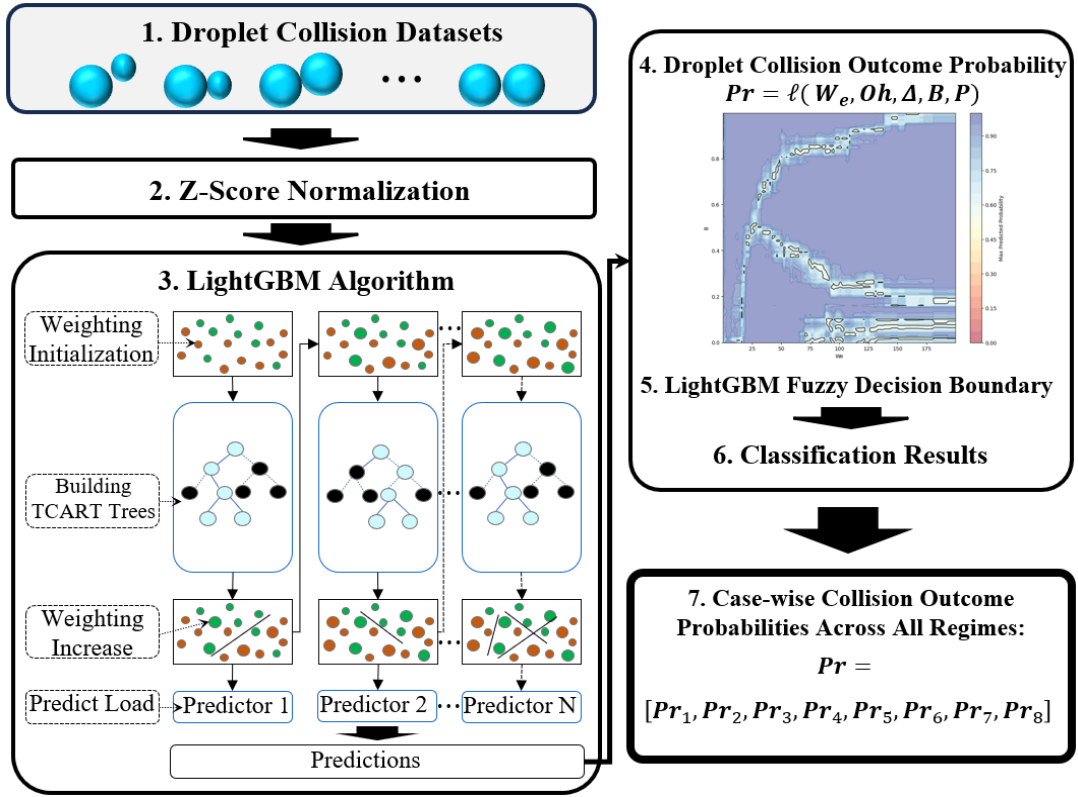

Fig 5. The machine learning workflow of lightGBM for predicting droplet collision outcomes.

**2.2.2 Logistic Regression Modeling for Model Application**

While the probabilistic outputs of LightGBM provide a flexible and accurate description of nonlinear decision boundaries, their tree-based representation remains implicit and difficult to interpret in reduced parameter subspaces. To obtain an explicit analytical form of the fuzzy boundaries, the probability fields learned by LightGBM are subsequently projected onto a multinomial logistic regression model. This step enables an effective polynomial approximation of the high-dimensional classification surface, facilitating both visualization and physical interpretation.

As shown in Fig. 6, the input parameter vector $x_i = [We_i, P_i, \Delta_i, B_i, Oh_i] \in \mathbb{R}^5$ is first expanded by a polynomial basis of degree two, including interaction terms among the collision parameters. This yields an augmented feature space,

$$\phi(x_i) = [We_i, \cdots, Oh_i, We_i^2, We_i P_i, \cdots, Oh_i^2] \tag{2}$$

which captures both quadratic effects and pairwise interactions that are known to influence droplet collision outcomes.

The multinomial logistic regression then models the probability of outcome class $k$ as

$$Pr_{i,k} = \frac{e^{(W_k^T \phi(x_i) + b_k)}}{\sum_{j=1}^{K} e^{(W_j^T \phi(x_i) + b_j)}} \tag{3}$$

where $W$ and $b$ denote the coefficient vector and bia for each class, respectively. The model parameters are obtained by maximizing the multinomial log-likelihood (equivalently, minimizing the cross-entropy loss):

$$L(y_i, \hat{y}_i) = -\sum_{i=1}^{N} \sum_{k=1}^{K} \mathbb{I}[y_i = k] \log Pr_{i,k} \tag{4}$$

with class balancing to account for the imbalanced distribution of collision outcomes.

By casting the classification in this probabilistic form, the logistic regression provides a principled means of linking the discrete droplet collision outcomes with continuous functions of the underlying parameters. This transformation reduces the complex, tree-based predictions to provide a mathematically tractable and interpretable description, thereby preserving the fuzzy nature of regime transitions. This representation is particularly valuable for systematic assessment of how impact parameters influence regime probabilities and for subsequent integration into the biased-dice sampling and broader physical modeling framework.

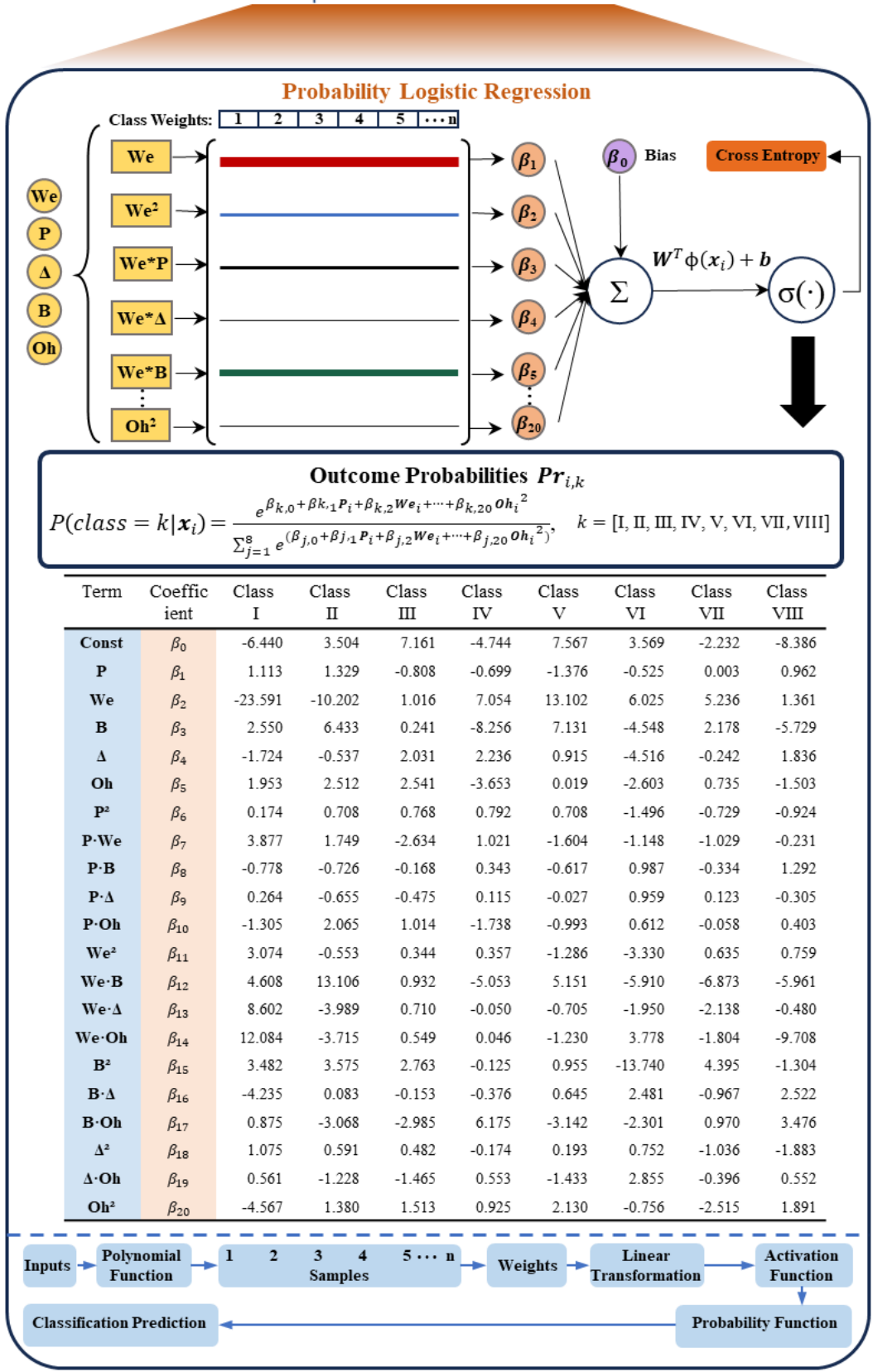

Fig 6. Probabilistic logistic regression for a droplet collision model learning from the trained machine learning model.

**2.2.3 Stochastic Classification Based on Logistic-Regression Probabilities**

Building upon the probabilistic classification obtained from multinomial logistic regression, we adopted a stochastic selection mechanism to process the fuzzy boundaries inherent to droplet collision regimes. In high-fidelity spray simulations, it is often necessary to produce a definite outcome for each realization. A naive strategy that selects the class with the highest probability disregards the stochastic nature of transitional boundaries, thereby introducing artificial determinism that can distort the physics of the modeled process.

As shown in Fig. 7, this approach treats the predicted probabilities of logistic regression as weights in a biased random sampling process. For a new sample $x_i$, the regime classification $C_i$ is obtained by sampling from the discrete distribution defined by

$$P_i = [Pr_{i,1}, Pr_{i,2}, Pr_{i,3}, Pr_{i,4}, Pr_{i,5}, Pr_{i,6}, Pr_{i,7}, Pr_{i,8}] \tag{4}$$

This biased dice mechanism transforms the probabilistic output of the logistic regression into a single categorical decision by drawing from an eight-class multinomial distribution

$$Pr(Y_1 = m_1, \cdots, Y_k = m_k) = \frac{n!}{m_1! m_2! \cdots m_k!} \prod_{k=1}^{K}(Pr_{i,k})^{m_k}, \sum_{k=1}^{K} m_k = n \tag{5}$$

where $m_k$ denotes the number of occurrences of class $k$ across $n$ trails.

By introducing this probabilistic sampling step, the approach preserves the inherent fuzziness at regime boundaries while yielding a definite class label for each realization. Therefore, this approach is particularly advantageous in engineering applications such as droplet-resolved simulations, where deterministic outcomes are required at the sample level but direct maximum-probability assignment would suppress physically meaningful variability.

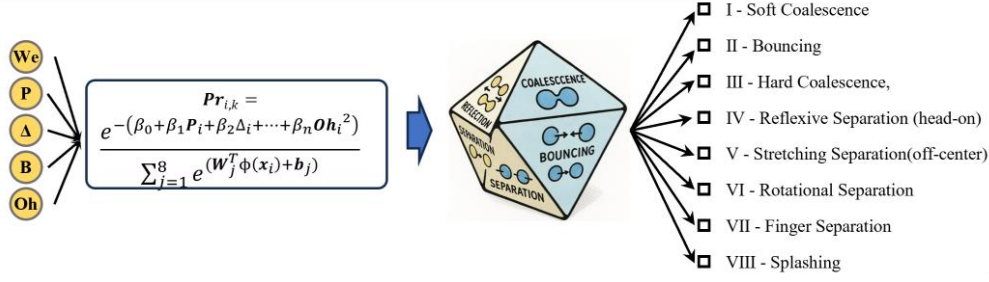

Fig 7. Stochastic classification of regime classification from eight-class logistic-regression probability outcomes using multinomial distribution sampling.

**2.3 Classification Performance Evaluation**

In the present methodology, the output of each stage represents a distinct form of classification, whether the raw probabilistic predictions from LightGBM, the analytically projected multinomial probabilities from logistic regression, or the stochastic outcomes derived from biased sampling. Therefore, unified performance evaluation matrices are required to quantify both the accuracy and the reliability of regime identification.

The confusion matrix $C$ is employed to provide a detailed description of classification errors and enable performance assessment. Each element $C_{ij}$ records the number of samples belonging to true class and predicted class, thereby representing true positives $TP_k = C_{kk}$, false positives $FP_k = \sum_{i \neq k} C_{ik}$, false negatives $FN_k = \sum_{j \neq k} C_{kj}$, and true negatives $TN_k = \sum_{i \neq k} \sum_{j \neq k} C_{ij}$. Based on these definitions, three performance indices are evaluated. Accuracy (AC) measures the overall correctness of classification

$$\text{Accuracy: } AC_k = \frac{TP_k + TN_k}{TP_k + FP_k + FN_k + TN_k} \tag{6}$$

However, global accuracy alone may mask systematic biases across outcome regimes. To address this, recall and specificity are introduced as

$$Recall_k = \frac{TP_k}{TP_k + FN_k}, Specificity_k = \frac{TN_k}{FP_k + TN_k} \tag{7}$$

Here, recall quantifies the sensitivity of the classifier to regime $k$, i.e. its ability to detect events of that type without omission, while specificity quantifies its reliability in excluding non-$k$ outcomes. To account for imbalance across regimes, we compute macro-averaged metrics as

$$Macro - M = \frac{1}{K}\sum_{k=1}^{K} M_k \tag{8}$$

Together, these metrics are particularly significant in the task of droplet collision outcome prediction, where imbalanced class distributions and fuzzy boundaries require balanced evaluation of both detection sensitivity and avoidance of spurious classification.

## 3. Results and Discussion

### 3.1 LightGBM Classification of Collision Outcomes

Figure 8(a) illustrates the classification results in representative two-dimensional parameter subspaces, which closely reproduces the structure of the raw data in Fig. 2, indicating that the classifier has captured the essential organizing features rather than imposing artificial boundaries. The confusion matrix in Fig. 8(b) provides a quantitative measure of classification accuracy. The strong diagonal dominance demonstrates that most regimes are sharply discriminated, with only limited overlap along known physical transitions. Misclassifications are rare and occur primarily along regime boundaries, particularly between regimes II and III, reflecting the intrinsic stochasticity of droplet interactions near critical thresholds.

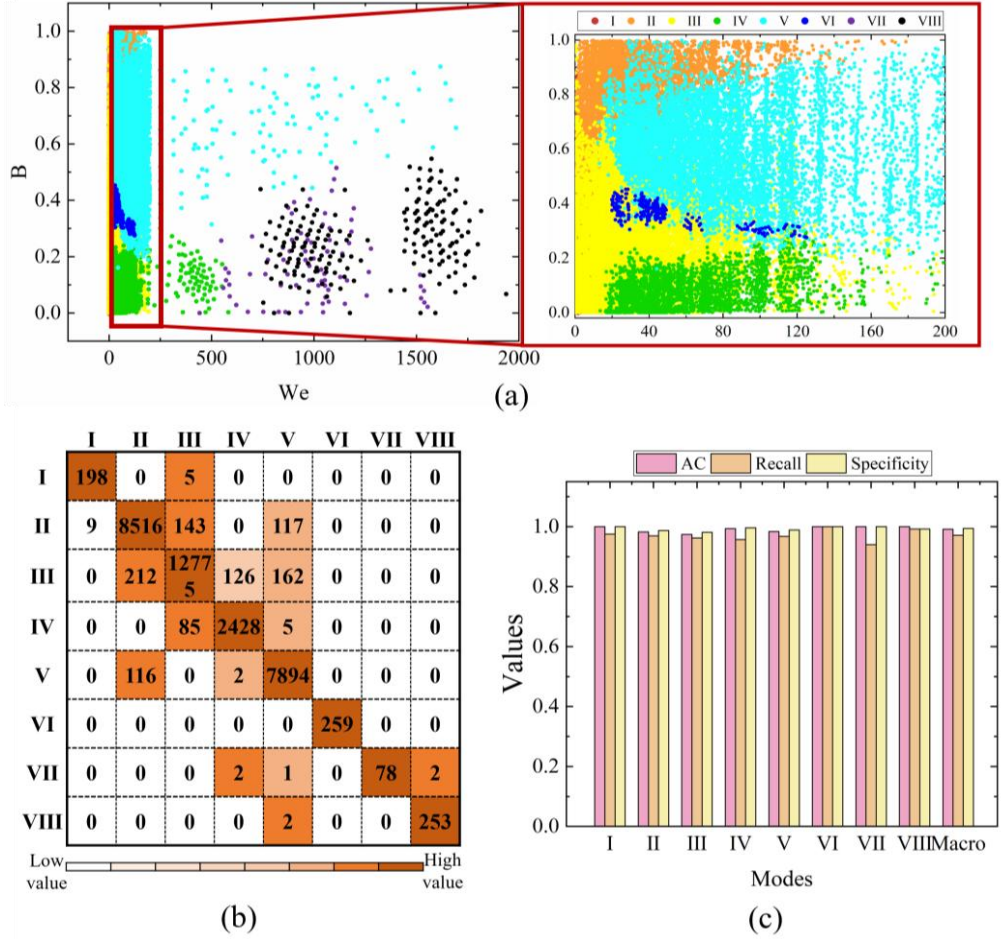

Fig 8. LightGBM classification performance: (a) Classification results, (b) Confusion matrix, and (c) Classification performance by modes.

The quantitative metrics in Fig. 8(c) confirm this overall picture. The macro-averaged accuracy of 0.9916, recall of 0.9717, and specificity of 0.9941 demonstrate both high overall correctness and strong robustness against false alarms. At the same time, the class-wise scores remain uniformly high across all regimes, showing that even categories represented by fewer samples are identified with the same level of confidence. To assess the robustness of the proposed LightGBM classifier, a 10-fold cross-validation was conducted to quantify the variation in classification performance under different training–testing splits in the Supplemental Material. The consistently low variance across folds indicates that the LightGBM model maintains stable classification capability.

Notably, these results demonstrate that LightGBM provides a highly reliable first-stage classifier, capable of mapping the complex regime boundaries with minimal error.

This capability is essential for the subsequent stages, where probabilistic formulations will be introduced to explicitly account for the fuzziness of boundaries.

**3.2 High-Dimensional Fuzzy Decision Boundaries of Regime Nomogram**

Figure 9 presents the high-dimensional fuzzy decision boundaries projected into representative two-dimensional subspaces, exemplified by the $We - B$ plane. A $100 \times 100$ grid was sampled over the $We - B$ range to evaluate ensemble predictions of the maximum class probability. The color contours illustrate the probabilistic landscape, with intermediate levels indicating gradual transitions between collision regimes. The contours reveal clear overlap between certain regimes, notably between regimes II and III, consistent with physical behaviors of droplet collisions near regime thresholds, where outcome uncertainty within these regimes is high due to subtle variations in collision conditions.

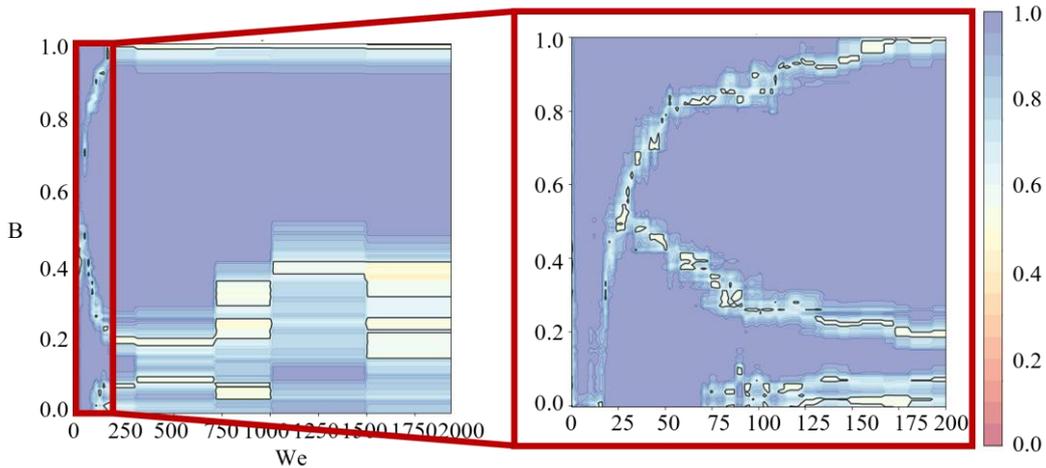

Fig 9. High-dimensional fuzzy decision boundares in $We - B$ regime nomogram.

The fuzzy decision boundary thus provides a richer, physically faithful depiction of regime structure, retaining classification confidence across the parameter space. This probabilistic boundary forms a key foundation for subsequent steps, enabling a stochastic treatment of regime determination and a more physically consistent representation of uncertainty.

**3.3 Probability Multinomial Logistic Regression of Fuzzy Decision Boundaries**

Figure 10 summarises the classification performance of the multinomial logistic regression. Figure 10(a) and (b) shows the classification results in representative two-dimensional subspaces, where the logistic regression preserves the general structure captured by LightGBM (Fig. 8), including the overall separation between regimes and the clustering of samples within each regime. This indicates that the regression model successfully distills the high-dimensional information encoded in the ensemble of decision trees into a simpler functional form.

The confusion matrix in Figure 10(c) further supports this observation. While the logistic regression yields slightly more off-diagonal entries compared to the LightGBM results (Fig. 8), these misclassifications remain largely confined to physically plausible boundaries—notably in regimes II and III—reflecting the intrinsic stochasticity of droplet interactions. This demonstrates that, despite the simplicity of the regression form, the model retains a high degree of fidelity to the complex decision boundaries learned by LightGBM.

Figure 10(d) quantifies these findings. Class-wise accuracy remains high across regimes, with even less represented regimes achieving robust recall and specificity. The macro-averaged accuracy of 93.2%, recall of 86.6%, and specificity of 96.9% are only modestly lower than those obtained with LightGBM (accuracy of 99.2%, recall of 97.2%, specificity of 99.4%), confirming that the regression model captures the core discriminative structure with minimal loss of performance. This comparison highlights the advantage of the proposed approach: it delivers a physically interpretable fuzzy decision boundary while closely approximating the classification capability of a far more complex ensemble model.

A multinomial logistic regression distills the complex decision structure learned by LightGBM into a compact, interpretable analytical form. This transformation not only facilitates efficient computation for subsequent stages but also provides explicit polynomial decision equations that can be directly inspected, thereby enhancing model transparency without sacrificing classification fidelity.

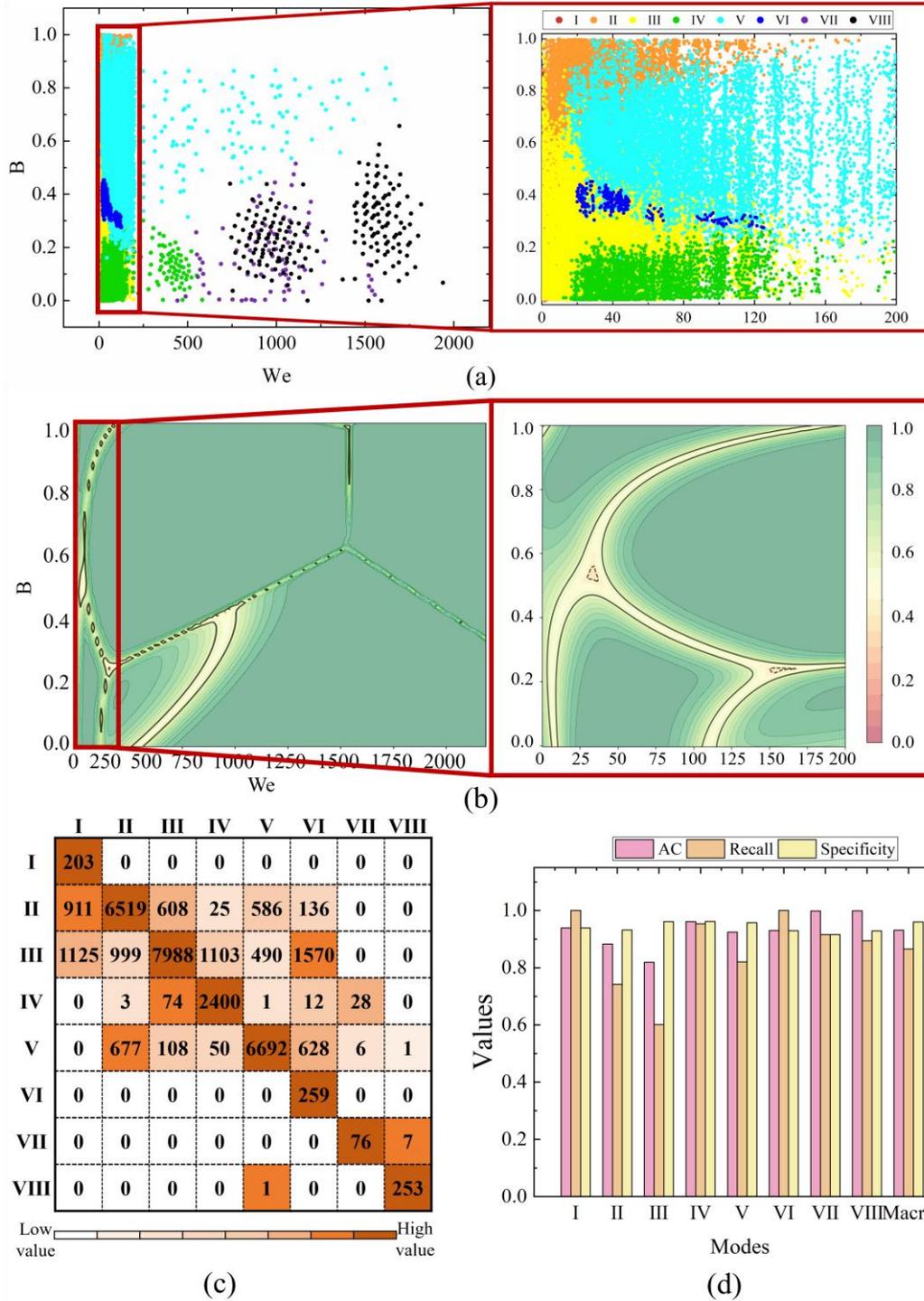

Fig 10. Logistic regression model classification performance: (a) Classification results, (b) High dimensional fuzzy decision boundary, (c) Confusion matrix, and (d) Classification performance.

### 3.4 Stochastic Realization of Collision Regime Probability Classification

To quantify the behavior of this biased-dice sampling, we performed 30 independent stochastic realizations and summarized per-class means and standard deviations in Fig. 11. The results in Fig. 11(a) show consistently high accuracy and extremely low standard deviations for all regimes, with most classes exceeding 0.94 mean values, indicating stable performance under repeated sampling. Specificity is also uniformly high for all classes ($> 0.90$), demonstrating that false positives are effectively suppressed across regimes. This consistency implies that the stochastic classification reliably preserves the integrity of regime identification, even under repeated realizations. Figure 11(b) summarises the macro-averaged performance over all classes. Accuracy and specificity remain consistently high across all classes, indicating robust classification and effective control of false positives. Recall shows slight variation, with several classes exhibiting lower values due to intrinsic overlaps in regime boundaries. This behavior is expected given the physical ambiguity in transitional droplet collisions. Nevertheless, high accuracy and specificity across all classes indicate that such misclassifications are largely confined to these boundary regions, underscoring the method's robustness in preserving physically meaningful classifications.

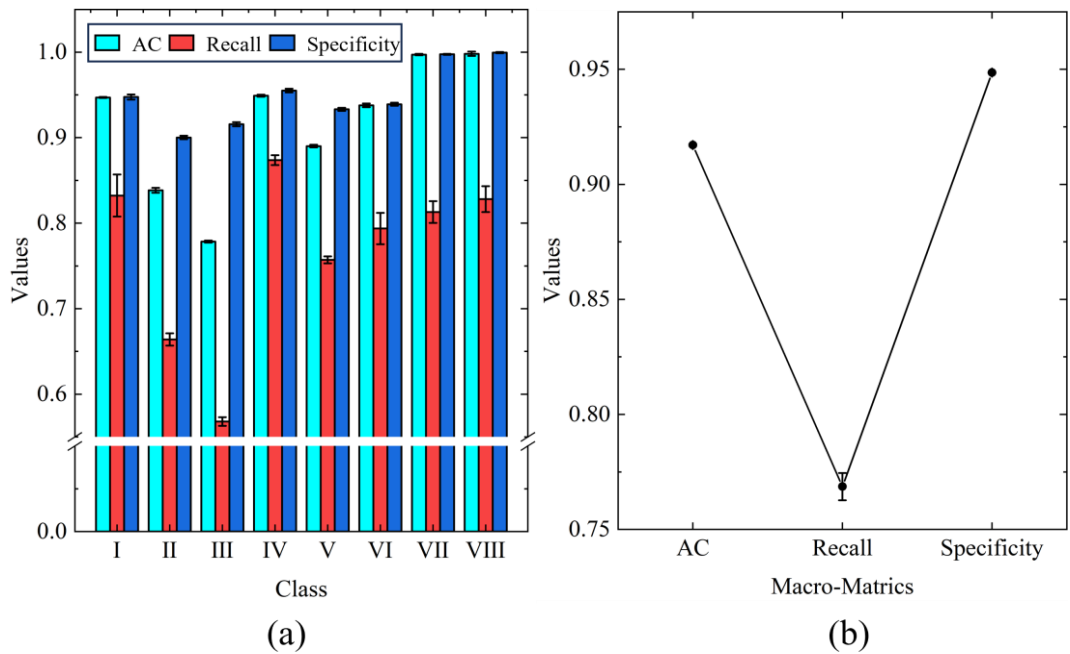

Fig 11. Stochastic classification performance: (a) Classification performance by modes, (b) Macro-M results.

## 4. Conclusions

This study established a practical, simulation-ready probabilistic data-driven model that quantitatively predicts binary droplet collision outcomes across wide parametric ranges. Each collision regime is represented by an explicit probability distribution rather than a fixed boundary, providing a continuous and physically consistent description of coalescence, bouncing, separation, and reflexive behaviors.

Trained on a comprehensive experimental dataset covering wide ranges of $We$, $Oh$, $B$, $\Delta$, and $\lambda$, the LightGBM and multinomial logistic regression-based learning approach can be directly used to compute regime probabilities under arbitrary input conditions, offering a reliable and interpretable tool for engineering and simulation applications. A biased-dice sampling (multinomial draws) according to the predicted eight-class probabilities further enables stochastic realizations of individual collision outcome prediction, preserving the natural variability and transitional uncertainty of droplet interactions while producing physically consistent outcomes.

Overall, the developed method delivers a comprehensive, probabilistic, and physically interpretable representation of binary droplet collisions. It effectively bridges the gap between experimental observation and computational modeling, enabling direct integration into large-scale flow simulations and establishing a foundation for uncertainty-aware predictions of droplet dynamics. Moreover, the framework can, in principle, accommodate additional collision parameters or higher-dimensional inputs, with practical limits set mainly by the availability and coverage of experimental data. Its tree-based LightGBM classifier and probabilistic sampling naturally handle multi-dimensional features without imposing constraints on the number or type of input variables. We recognize that the model's performance is constrained by the scope of its training data, which is derived from previously non-unified experiments. Specifically, the model is trained on experimental data in the wide ranges of ranges of $We = 0 \sim 2000$, $B = 0 \sim 1.0$, $Oh = 2.7 \times 10^{-3} \sim 5.5 \times 10^{-1}$, $P = 0.6 \sim 9.0$, and $\Delta = 1.0 - 5.0$, clarifying the scope of applicability and limits of extrapolation. Its predictive capability may therefore diminish for extreme conditions outside these trained parameter ranges.

Future efforts will focus on evolving the model from a classifier into a comprehensive "digital twin" by developing companion models that quantify detailed collision outcomes. The proposed probabilistic model can also be directly incorporated into Eulerian–Lagrangian spray simulation frameworks; a schematic illustration and pseudocode for practical integration into CFD codes are provided in the Supplementary Material. The seamless integration of this model into high-fidelity spray simulation codes and its rigorous validation against macroscopic spray characteristics constitute the crucial next steps to fully realize its potential.

**Acknowledgements.** P.Z. acknowledges support from the National Natural Science Foundation of China (No. 52176134) and partially from the APRC-CityU New Research Initiatives/Infrastructure Support from Central of City University of Hong Kong (No. 9610601). The authors are grateful to the National Supercomputer Center in Guangzhou (Tianhe-2) for supporting the GPU computing.

**Declaration of interests.** The authors report no conflict of interest.

**Data availability.** The data that support the findings of this study are available from the corresponding author upon reasonable request.

**Supplementary material.** Supporting information of the present neural network and numerical simulations is available.

**References**
G. M. McFarquhar, A new representation of collision-induced breakup of raindrops and its implications for the shapes of raindrop size distributions, *Journal of the Atmospheric Sciences*, vol. 61, no. 7, pp. 777-794, 2004.
A. van der Bos, et al., Velocity profile inside piezoacoustic inkjet droplets in flight: comparison between experiment and numerical simulation, *Physical review applied*, vol. 1, no. 1, pp. 014004, 2014.
B. Kichatov, et al., Droplet manipulation in liquid flow using of magnetic micromotors for drug delivery and microfluidic systems, *Colloids and Surfaces A: Physicochemical and Engineering Aspects*, vol. 691, pp. 133891, 2024.
W. A. Sirignano, Fuel droplet vaporization and spray combustion theory, *Prog. Energy Combust. Sci.*, vol. 9, no. 4, pp. 291-322, 1983.
S. Kim, et al., Modeling of binary droplet collisions for application to inter-impingement sprays, *Int. J. Multiph. Flow*, vol. 35, no. 6, pp. 533-549, 2009.
I. Makhnenko, et al., A review of liquid sheet breakup: Perspectives from agricultural sprays, *Journal of Aerosol Science*, vol. 157, pp. 105805, 2021.
Z. Ren, et al., Supersonic spray combustion subject to scramjets: Progress and challenges, *Progress in Aerospace Sciences*, vol. 105, pp. 40-59, 2019.

Z. An, et al., Recent progresses in research on liquid ammonia spray and combustion: A review, *Applications in Energy and Combustion Science*, vol. 20, pp. 100293, 2024.

X. Li, et al., A review on the recent advances of flash boiling atomization and combustion applications, *Prog. Energy Combust. Sci.*, vol. 100, pp. 101119, 2024.

Z. Hu, The role of raindrop coalescence and breakup in rainfall modeling, *Atmospheric research*, vol. 37, no. 4, pp. 343-359, 1995.

N. Mishra, et al., Nano emulsion drug delivery system: A review, *Current Nanomedicine (Formerly: Recent Patents on Nanomedicine)*, vol. 13, no. 1, pp. 2-16, 2023.

S. Lain, M. Sommerfeld, Influence of droplet collision modelling in Euler/Lagrange calculations of spray evolution, *Int. J. Multiph. Flow*, vol. 132, pp. 103392, 2020.

S. S. Sazhin, *Droplets and sprays: Simple models of complex processes*, Springer, 2022.

Y. Jiang, et al., An experimental investigation on the collision behaviour of hydrocarbon droplets, *J. Fluid Mech.*, vol. 234, pp. 171-190, 1992.

J. Qian, C. K. Law, Regimes of coalescence and separation in droplet collision, *J. Fluid Mech.*, vol. 331, pp. 59-80, 1997.

K.-L. Pan, et al., Binary droplet collision at high Weber number, *Physical Review E — Statistical, Nonlinear, and Soft Matter Physics*, vol. 80, no. 3, pp. 036301, 2009.

C. Tang, et al., Bouncing, coalescence, and separation in head-on collision of unequal-size droplets, *Phys. Fluids*, vol. 24, no. 2, pp. 2012.

K.-L. Huang, et al., Pinching dynamics and satellite droplet formation in symmetrical droplet collisions, *Phys. Rev. Lett.*, vol. 123, no. 23, pp. 234502, 2019.

K.-L. Pan, et al., Rotational separation after temporary coalescence in binary droplet collisions, *Phys. Rev. Fluids*, vol. 4, no. 12, pp. 123602, 2019.

K.-L. Huang, K.-L. Pan, Transitions of bouncing and coalescence in binary droplet collisions, *J. Fluid Mech.*, vol. 928, pp. A7, 2021.

X. Xia, et al., Scaling law in the inviscid coalescence of unequal-size droplets, *J. Fluid Mech.*, vol. 1010, pp. R2, 2025.

D. Zhou, et al., Intense deformation and fragmentation of two droplet collision at high Weber numbers, *Colloids and Surfaces A: Physicochemical and Engineering Aspects*, vol. 655, pp. 130171, 2022.

K. H. Al-Dirawi, A. E. Bayly, A new model for the bouncing regime boundary in binary droplet collisions, *Phys. Fluids*, vol. 31, no. 2, pp. 2019.

M. Sui, et al., Extended model of bouncing boundary for droplet collisions considering numerous different liquids, *Int. J. Multiph. Flow*, vol. 162, pp. 104418, 2023.

M. Sommerfeld, M. Kuschel, Modelling droplet collision outcomes for different substances and viscosities, *Exp. Fluids*, vol. 57, no. 12, pp. 187, 2016.

N. Ashgriz, J. Poo, Coalescence and separation in binary collisions of liquid drops, *J. Fluid Mech.*, vol. 221, pp. 183-204, 1990.

S. Suo, M. Jia, Correction and improvement of a widely used droplet–droplet collision outcome model, *Phys. Fluids*, vol. 32, no. 11, pp. 2020.

T. L. Georjon, R. D. Reitz, A drop-shattering collision model for multidimensional spray computations, *At. Sprays*, vol. 9, no. 3, pp. 1999.

S. L. Post, J. Abraham, Modeling the outcome of drop–drop collisions in Diesel sprays, *Int. J. Multiph. Flow*, vol. 28, no. 6, pp. 997-1019, 2002.

G. Luret, et al., Modeling collision outcome in moderately dense sprays, *At. Sprays*, vol. 20, no. 3, pp. 2010.

V. Tyurenkova, et al., Mathematical modeling of droplet collisions in sprays under microgravity conditions, *Acta Astronautica*, vol. 219, pp. 459-466, 2024.

J.-P. Estrade, et al., Experimental investigation of dynamic binary collision of ethanol droplets– a model for droplet coalescence and bouncing, *International Journal of Heat and Fluid Flow*, vol. 20, no. 5, pp. 486-491, 1999.

P. Brazier-Smith, et al., The interaction of falling water drops: coalescence, *Proceedings of the Royal Society of London. A. Mathematical and Physical Sciences*, vol. 326, no. 1566, pp. 393-408, 1972.

A. Agarwal, et al., The computational cost and accuracy of spray droplet collision models, SAE Technical Paper, Tech. Rep. 0148-7191, 2019.

W. Yu, S. Chang, A machine learning-based approach to predict the outcome of binary droplet collision, *Chemical Engineering Science*, vol. 319, no. pp. 122349, 2026.

A. Munnannur, R. D. Reitz. Droplet collision modeling in multi-dimensional spray computations, *SAE World Congress*, 2007.

L. Au-Yeung, P. A. Tsai, Predicting impact outcomes and maximum spreading of drop impact on heated nanostructures using machine learning, *Langmuir*, vol. 39, no. 50, pp. 18327-18341, 2023.

J. Tang, et al., Universal model for predicting maximum spreading of drop impact on a smooth surface developed using boosting machine learning models, *Industrial & Engineering Chemistry Research*, vol. 62, no. 37, pp. 15268-15277, 2023.

J. Yee, et al., Prediction of the morphological evolution of a splashing drop using an encoder– decoder, *Machine Learning: Science and Technology*, vol. 4, no. 2, pp. 025002, 2023.

M. Pierzyna, et al., Data-driven splashing threshold model for drop impact on dry smooth surfaces, *Phys. Fluids*, vol. 33, no. 12, pp. 2021.

H. Ye, et al., Machine learning-based splash prediction model for drops impact on dry solid surfaces, *Phys. Fluids*, vol. 35, no. 9, pp. 2023.

A. Agarwal, Machine learning models for prediction of droplet collision outcomes, *arXiv preprint arXiv:2110.00167*, pp. 2021.

D. A. de Aguiar, et al., Predicting energy budgets in droplet dynamics: A recurrent neural network approach, *International Journal for Numerical Methods in Fluids*, vol. 97, no. 5, pp. 854-873, 2025.

D. Kochkov, et al., Machine learning–accelerated computational fluid dynamics, *Proc. Natl. Acad. Sci. U.S.A.*, vol. 118, no. 21, pp. e2101784118, 2021.

F. Salehi, et al., Data-driven modelling of spray flows: Current status and future direction, *Journal of the Energy Institute*, vol., pp. 101991, 2025.

C. Hu, et al., Three-dimensional numerical investigation and modeling of binary alumina droplet collisions, *International Journal of Heat and Mass Transfer*, vol. 113, pp. 569-588, 2017.

M. Sui, et al. Modelling the occurrence of bouncing in droplet collision for different liquids, *Proceedings of the ILASS2019-29th European Conference on Liquid Atomization and Spray Systems, Paris, France, 2-4 September 2019*, Paris, France, 2019.

J. H. Evans, Dimensional analysis and the Buckingham Pi theorem, *American Journal of Physics*, vol. 40, no. 12, pp. 1815-1822, 1972.

G. Brenn, et al., The formation of satellite droplets by unstable binary drop collisions, *Phys. Fluids*, vol. 13, no. 9, pp. 2463-2477, 2001.

G. Brenn, V. Kolobaric, Satellite droplet formation by unstable binary drop collisions, *Phys. Fluids*, vol. 18, no. 8, pp. 2006.

C. Rabe, et al., Experimental investigation of water droplet binary collisions and description of outcomes with a symmetric Weber number, *Phys. Fluids*, vol. 22, no. 4, pp. 2010.

C. Planchette, et al. Binary collisions of immiscible liquid drops for liquid encapsulation, *7th Int. Conf. Multiphase Flow (ICMF 2010)*, Place, 9.7. 4.--, Year.

A. Foissac, et al. Binary water droplet collision study in presence of solid aerosols in air, *Proc. 7th Int. Conf. Multiphase Flow (ICMF)*, 2010.

M. Kuschel, M. Sommerfeld, Investigation of droplet collisions for solutions with different solids content, *Exp. Fluids*, vol. 54, no. 2, pp. 1440, 2013.

H. Hinterbichler, et al., Ternary drop collisions, *Exp. Fluids*, vol. 56, no. 10, pp. 190, 2015.

K.-L. Pan, et al., Controlling droplet bouncing and coalescence with surfactant, *J. Fluid Mech.*, vol. 799, pp. 603-636, 2016.

L. Reitter, et al. Experimental and computational investigation of binary drop collisions under elevated pressure, *Ilass Europe. 28th european conference on Liquid Atomization and Spray Systems*, Editorial Universitat Politècnica de València, 815-821, 2017.

G. Finotello, et al., The dynamics of milk droplet–droplet collisions, *Exp. Fluids*, vol. 59, no. 1, pp. 17, 2018.

M. Sommerfeld, L. Pasternak, Advances in modelling of binary droplet collision outcomes in sprays: a review of available knowledge, *Int. J. Multiph. Flow*, vol. 117, pp. 182-205, 2019.

M. Sommerfeld, L. Pasternak. Experimental studies on binary water droplet collisions considering size ratio effects, *International Conference on Liquid Atomization and Spray Systems (ICLASS)*, Place, 1, 2021.

L. P. McCarthy, et al., Dynamics and outcomes of binary collisions of equi-diameter picolitre droplets with identical viscosities, *Physical Chemistry Chemical Physics*, vol. 24, no. 35, pp. 21242-21249, 2022.

D. Baumgartner, et al., Universality of stretching separation, *J. Fluid Mech.*, vol. 937, pp. R1, 2022.

G. Ke, et al., Lightgbm: A highly efficient gradient boosting decision tree, *Advances in neural information processing systems*, vol. 30, pp. 2017.

H. Henderi, et al., Comparison of Min-Max normalization and Z-Score Normalization in the K-nearest neighbor (kNN) Algorithm to Test the Accuracy of Types of Breast Cancer, *International Journal of Informatics and Information Systems*, vol. 4, no. 1, pp. 13-20, 2021.